\documentclass[twocolumn,preprintnumbers,amsmath,amssymb,prl]{revtex4}
\usepackage{pifont}
\usepackage{amsmath}
\usepackage{amsfonts}
\usepackage{graphicx}
\usepackage{amssymb}
\usepackage{dcolumn}
\usepackage{bm}
\usepackage{slashed}

\begin{document}

\title{Analytical solution for Klein-Gordon equation and action function of the solution for Dirac equation in counter-propagating laser waves }

\author{Huayu Hu$^{\ast}$}
\author{Jie Huang}

\affiliation{Hypervelocity Aerodynamics Institute, China Aerodynamics Research and Development Center, 621000 Mianyang, Sichuan, China}

\date{\today}

\begin{abstract}
Nonperturbative calculation of QED processes participated by a strong electromagnetic field, especially provided by strong laser facilities at present and in the near future, generally resorts to the Furry picture with the usage of analytical solutions of the particle dynamical equation, such as the Klein-Gordon equation and Dirac equation. However only for limited field configurations such as a plane-wave field could the equations be solved analytically. Studies have shown significant interests in QED processes in a strong field composed of two counter-propagating laser waves, but the exact solutions in such a field is out of reach. In this paper, inspired by the observation of the structure of the solutions in a plane-wave field, we develop a new method and obtain the analytical solution for the Klein-Gordon equation and equivalently the action function of the solution for the Dirac equation in this field, under a largest dynamical parameter condition that there exists an inertial frame in which the particle free momentum is far larger than the other field dynamical parameters. The applicable range of the new solution is demonstrated and its validity is proven clearly. The result has the advantage of Lorentz covariance, clear structure and close similarity to the solution in a plane-wave field, and thus favors convenient application.
\end{abstract}


\maketitle

\section{I. Introduction}
QED is the successful theory to describe the interaction between particles and photons. Conventionally the calculation is carried out in a perturbative manner since the interaction is characterized by the fine-structure constant $\alpha\ll 1$.  However, in an intense laser field with $A_\mu$ being the four potential, the laser-particle interaction is characterized by the classical nonlinearity parameter $\xi=\frac{\sqrt{-<e^2A^2>}}{m_e}$ where $e$ is the electron charge, $m_e$ is the electron mass and $<>$ represents time averaging \cite{Ritus}. If $\xi\gtrsim1$ the interaction is in the nonperturbative multi-photon regime. For example, the SLAC E-144 experiment with $\xi\approx0.3$ found a electron-positron-pair-production rate scaling of $R\sim\xi^{10}$ \cite{SLAC} and in the perturbative theory this would be interpreted as the typical $\xi^{2N_0}$ dependence with $N_0=5$ photon absorption. However, the calculation \cite{H2010} reveals that on average more than 6 photons are absorbed in the process, and thus demonstrates the onset of nonperturbative effects on the experiment. It is also found by calculation that if the intensity of the laser field in the above experiment is enhanced that $\xi\sim 1$, photon orders up to $N\approx50$ give significant contributions to the total rate and thus the process enters the fully nonperturbative regime. Therefore, special techniques are required to tackle such strong-field nonperturbative problems in order to take the effects of the laser field properly into account.

The general approach is to employ the Furry picture \cite{Furry}, where the laser field is treated as a classical background field and the particle state is represented by the exact solution of the particle dynamical equation in the laser field, so called laser-dressed state. For the ideal situation of the laser field being a plane wave or a constant-crossed field, analytical solutions of the Klein-Gordon equation for a meson and the Dirac equation for a fermion have been obtained exactly \cite{Volkov, Reiss, BrownKibble, RitusCC, Hartin}. By using such laser-dressed state the particle-laser interaction has been taken into account to all orders, and the remaining interaction between the laser-dressed particle and the QED vacuum is weak, thus allowing calculations in a perturbative scheme similar to the conventional QED. Various strong-field QED processes in a plane-wave field (or an approximated plane-wave field with the spot radius of the laser beam being much larger than the laser central wavelength \cite{RMP2012}) have been investigated by this method, including laser-assisted bremsstrahlung \cite{mpBrem}, multi-photon Compton scattering \cite{BrownKibble, mpCompton, mpCompton1}, electron-positron pair production \cite{H2010, mpPP, mpPP1} and so on.

For an arbitrary non-plane-wave field, exact analytical solutions for the above equations are generally out of reach. However, several laser fields of particular interest in study are of this kind, e.g., the field composed of two counter-propagating laser waves. The colliding wave configuration can provide higher field intensity and the electron classical trajectory is distinctly different compared to that in the plane-wave field. Extensive studies suggest that this kind of laser field is efficient in producing electron-positron pairs and supporting avalanche pair production, and thus is ideal for vacuum cascade observation \cite{Bell2008, Bulanov2010S, Bulanov2010, Kirk2009, Nerush2011}. Without analytical Volkov solution of Dirac equation in this field, the pair production rate has been calculated under the important quasi-stationarity approximation \cite{Kirk2009}: the reaction rate for each pair production is calculated with the value of the external field fixed at the given time, and finally these rates are time-averaged over the total interaction time. This approximation is justified if $\xi\gg 1$ since the formation length of a QED process in a plane-wave field is $\xi$ times smaller than the laser wavelength \cite{Sebastian, Ritus}. In \cite{Hu2014} a complete QED calculation based on Volkov solution in a plane-wave field is carried out for the electron-positron pair production in the impact of a 50MeV electron with two colliding 10keV X-ray laser beams each of the intensity $10^{20}$W/cm$^2$. This treatment takes the advantage that $\xi\ll1$ though the absolute intensity is high. Although it is by this measure not a strong-field problem, novel features in the pair production process compared to the plane-wave case are identified. Therefore, it is meaningful to investigate analytical solutions for the particle dynamical equations in non-plane-wave fields to calculate more accurate strong-field QED reaction rates in these fields, especially in the regime with intermediate values of $\xi$. The recent breakthrough in solving the Dirac equation in propagating laser waves is worked out by A. Di Piazza in \cite{PiazzaG, PiazzaG2}. Applying the WKB method and looking for a solution of the form $\psi(x)=\exp[iS(x)/\hbar]\varphi(x)$, the action function $S(x)$ is derived first by solving the classical electron dynamical equation in the background electromagnetic field and then the bi-spinor $\varphi(x)$ is constructed via the method of characteristics.  In this way the electron wave functions in the presence of a background electromagnetic field of a general space-time structure are constructed in particular inertial frames where the initial energy of the electron is the largest dynamical energy scale. In \cite{PiazzaG2} it is argued that even in view of the future strong laser facilities, such as the Extreme Light Infrastructure (ELI) \cite{ELI} and the Exawatt Center for Extreme Light Studies (XCELS) \cite{XCELS}, it is still necessary to employ ultrarelativistic electrons in experimental studies on strong-field QED problems in the quantum nonlinearity regime where not only $\xi\gtrsim 1$ but also the quantum nonlinearity parameter $\chi=\xi\frac{\omega_b}{m_e}\gtrsim 1$ with $\omega_b$ being the photon energy in the rest frame of the electron.

In this paper, we develop a new method and analytically solve the Klein-Gordon equation in a background electromagnetic field composed of two counter-propagating laser waves in a Lorentz covariant manner. The method is inspired by the observation that if the coefficients of the Fourier expansion of the solution in a plane-wave field to different photon modes are written in the form of a Bessel function, all  parameters can be determined by simple rules, as demonstrated in section 2. For the solution in the non-plane-wave field under study, the coefficients of the Fourier expansion of the solution to different modes of the two kinds of photons are written as a multiplication of Bessel functions, and the parameters are determined by rules in analogy or as a development to the simple rules. The applicable range is obtained by examining the validity of approximations used in derivation steps, and in this way a largest dynamical parameter condition is imposed, that there exists an inertial frame in which the particle free momentum is far larger than the other field dynamical parameters. We can have in mind the case of an energetic particle obliquely impacting on two counter-propagating optical or X-ray intense laser waves in the lab frame, while the calculation can be conducted in arbitrary inertial frames. This is the content of section 3. By solving the Klein-Gordon equation, the action function $S(x)$ of the solution of the Dirac equation is obtained, because if the solution $\psi=\exp[iS'(x)/\hbar]$ satisfies the Klein-Gordon equation
\begin{equation}\label{KG}
[(i\hbar\partial_\mu-eA_\mu)^2-m_e^2]\psi=0,
\end{equation}
then except for a term proportional to $\hbar$ it results in
\begin{equation}\label{action}
(\partial_\mu S'+eA_\mu)(\partial^\mu S'+eA^\mu)-m_e^2=0,
\end{equation}
which is just the equation determining the action function $S(x)$ for Dirac equation \cite{PiazzaG, PiazzaG2}.

In section 4 the newly obtained solution is justified by showing that solutions in a plane-wave field can be recovered naturally from it. In section 5 the solution is simplified and takes a form in close similarity to the solution in a plane-wave field. Finally it is validated by substituting the action function into the basic equation (\ref{action}) and proving the consistency.

In the following the natural units $c=\hbar=1$ is used unless claimed otherwise. In the context the electron charge $e$ should not be confused with the exponential constant in exponential expressions.

\section{II. Reconstruction of the solution for Klein-Gordon equation in a circularly polarized plane wave}
We consider the solution for the Klein-Gordon equation (\ref{KG}) in a circularly polarized plane wave $A=a_1\cos(k\cdot x)+a_2\sin(k\cdot x)$, of which the Fourier expansion to different photon modes takes the form
\begin{equation}\label{psiG}
\psi=e^{\pm i q_\pm\cdot x}\sum_{n=-\infty}^{\infty}c_n J_n(\alpha)e^{i n \beta}e^{i n k\cdot x},
\end{equation}
where the dressed momentum is $q_\pm=p_\pm+\frac{m_e^2\xi^2 }{2k\cdot p_\pm}k$ with $p_\pm$ being the particle momentum outside the electromagnetic field, and $J_n$ is the Bessel function. Our question is if we only know the form of the solution is given by Eq. (\ref{psiG}), how to determine the parameters $c_n$, $\alpha$ and $\beta$ so that $\psi$ can satisfy Eq. (\ref{KG})? The answer seems to be trivial, but the approach to actually obtain these parameters can provide useful information.

Substituting Eq. (\ref{psiG}) into Eq. (\ref{KG}), we obtain that
\begin{align}
\sum_{n=-\infty}^{\infty} c_n & J_n(\alpha) e^{i n \beta} [2 n k\cdot p_\pm+e p_\pm\cdot(a_1-i a_2)e^{i k\cdot x}\nonumber\\
&+e p_\pm \cdot(a_1+i a_2)e^{-i k\cdot x}] e^{i n k\cdot x}=0.
\end{align}
It results in the sequence of equations related to different photon modes
\begin{align}\label{psiG1}
c_n J_n&(\alpha)2 n k\cdot p_\pm+ c_{n-1} J_{n-1}(\alpha)e^{-i \beta}e p_\pm\cdot(a_1-i a_2)\nonumber\\
&+ c_{n+1} J_{n+1}(\alpha)e^{i \beta}e p_\pm\cdot(a_1+i a_2)=0.
\end{align}
Compared to the identity of the Bessel function
\begin{equation}\label{BR}
\frac{\alpha}{2n}(J_{n-1}(\alpha)+J_{n+1}(\alpha))=J_n(\alpha),
\end{equation}
it can be found that in Eq. (\ref{psiG1}) if we let
\begin{equation}\label{rule1}
 c_n=1,
\end{equation}
and
\begin{equation}\label{rule2}
\textrm{Im}[e^{-i \beta}e p_\pm\cdot(a_1-i a_2)]=0,
\end{equation}
the equation becomes
\begin{align}\label{psiG2}
\frac{-\textrm{Re}[e^{-i \beta}e p_\pm\cdot(a_1-i a_2)]}{2nk\cdot p_\pm}(J_{n-1}(\alpha)+J_{n+1}(\alpha))=J_n(\alpha),
\end{align}
and thus
\begin{equation}\label{rule3}
\alpha=-\frac{\textrm{Re}[e^{-i \beta}e p_\pm\cdot(a_1-i a_2)]}{k\cdot p_\pm}.
\end{equation}
From Eq. (\ref{rule2}) and Eq. (\ref{rule3}) $\alpha$ and $\beta$ can be written in the familiar form
\begin{align}
&\cos\beta=\frac{p_\pm\cdot a_1}{\sqrt{(p_\pm\cdot a_1)^2+(p_\pm\cdot a_2)^2}},\\
&\sin\beta=-\frac{p_\pm\cdot a_2}{\sqrt{(p_\pm\cdot a_1)^2+(p_\pm\cdot a_2)^2}},\\
&\alpha=-\frac{e\sqrt{(p_\pm\cdot a_1)^2+(p_\pm\cdot a_2)^2}}{k\cdot p_\pm}.\label{result1}
\end{align}
Another solution is got by changing the sign of the above quantities, corresponding to $\beta=\beta+\pi$ and $\alpha=-\alpha$, but this leads to the same wave function (\ref{psiG}) since $e^{i n(\beta+\pi)}J_n(-\alpha)=(-1)^{2n}e^{i n \beta}J_n(\alpha)$.

In this section the coefficients of the Fourier expansion of the solution for the Klein-Gordon equation are acquired in an empirical way, in which the identity (\ref{BR}) plays a crucial role. In the following it is shown that the experiences and observations gained here can be used to tackle nontrivial problems.

\section{III. Construction of the solution for Klein-Gordon equation in counter-propagating waves}\label{seccon}
Our aim is to find the solution for the Klein-Gordon equation and equivalently the action function for the Dirac equation in a non-plane-wave field. Consider
\begin{equation}\label{KG3}
[(\hat{p}-eA-eA')^2-m_e^2]\psi=0,
\end{equation}
where $A=a[\epsilon_1\cos(k\cdot x)+\epsilon_2\sin(k\cdot x)]$ and $A'=a'[\epsilon'_1\cos(k'\cdot x)+\epsilon'_2\sin(k'\cdot x)]$ respectively represent a circularly polarized plane wave with the non-plane-wave condition $k\cdot k'\neq 0$. As a special case, consider the two waves counter propagate with each other. Without loss of generality, suppose $k=(\omega,0,0,k_z)$ and $k'=(\omega',0,0,k'_z)$ where $k_z$ is positive and $k'_z$ is negative, and let $\epsilon_1=\epsilon'_1=(0,1,0,0)$ and $\epsilon_2=\epsilon'_2=(0,0,1,0)$.


Based on the observation of the structure of the solution (\ref{psiG}) in a plane-wave field, we assume the Fourier expansion of the solution here takes the form
\begin{equation}\label{psiG3}
\psi=e^{\pm i p'_\pm\cdot x}\sum_{n=-\infty}^{\infty}\sum_{m=-\infty}^{\infty}C_{nm}e^{i n k\cdot x}e^{i m k'\cdot x},
\end{equation}
where
\begin{equation}\label{Cnm}
C_{nm}=c_{nm} J_n(\alpha_m)e^{i n \beta}J_m(\alpha'_n)e^{i m \beta'},
\end{equation}
and the formal dressed momentum is written as
\begin{equation}
p'_\pm=p_\pm+\frac{e^2 a^2}{2k\cdot p_\pm}k+\frac{e^2 a'^2}{2k'\cdot p_\pm}k' .
\end{equation}
The arguments of the Bessel functions are assumed to be index dependent, the reason of which will be shown later. The quantity $p'_\pm$ is not the physical dressed momentum, which can only be determined after the acquisition of $C_{nm}$. The solution (\ref{psiG3}) would be obtained by determining $c_{nm}$, $\alpha_m$, $\alpha'_n$, $\beta$ and $\beta'$.

Substituting Eq. (\ref{psiG3}) into Eq. (\ref{KG3}), a sequence of equations related to different photon modes can be derived. Considering the possibility of $c_{nm}$ to be a function of the time-space coordinates, we get
\begin{align}\label{KGD3}
&c_{nm}J_n(\alpha_m)J_m(\alpha'_n)(r_0+r_1 n+r_2 m+r_3 nm)\nonumber\\
&+c_{n-1,m}J_{n-1}(\alpha_m)J_m(\alpha'_{n-1}) e^{-i\beta}b\nonumber\\
&+c_{n+1,m}J_{n+1}(\alpha_m)J_m(\alpha'_{n+1}) e^{i\beta}b^*\nonumber\\
&+c_{n,m-1}J_{n}(\alpha_{m-1})J_{m-1}(\alpha'_n) e^{-i\beta'}d\nonumber\\
&+c_{n,m+1}J_{n}(\alpha_{m+1})J_{m+1}(\alpha'_n) e^{i\beta'}d^*=0 .
\end{align}
with
\begin{align}\label{KGD3C2}
&r_0=\frac{e^4a^2a'^2k\cdot k'}{2(k\cdot p_\pm)(k'\cdot p_\pm)}+2e^2A\cdot A'\nonumber\\
&+\frac{(\hat{p}^2-2eA\cdot \hat{p}-2eA'\cdot \hat{p})c_{nm}}{c_{nm}}-\frac{2}{c_{nm}}(\hat{p}_\mu c_{nm})(\pm {p'}_\pm^{\mu}),\\
&r_1=\pm2k\cdot p_\pm\pm\frac{e^2a'^2}{k'\cdot p_\pm}k\cdot k'-\frac{2}{c_{nm}}(\hat{p}_\mu c_{nm})k^\mu,\\
&r_2=\pm2k'\cdot p_\pm\pm\frac{e^2a^2}{k\cdot p_\pm}k\cdot k'-\frac{2}{c_{nm}}(\hat{p}_\mu c_{nm})k'^\mu,\\
&r_3=2k\cdot k',\\
&b=\pm e a p_\pm\cdot(\epsilon_1-\textit{i} \epsilon_2),\\
&d=\pm e a' p_\pm\cdot(\epsilon'_1-\textit{i} \epsilon'_2).
\end{align}
It is found that, the empirical method (\ref{rule1}---\ref{rule3}) based on the identity (\ref{BR}) can be applied. Let
\begin{align}
&r_0=0,\label{r0}\\
&\textrm{Im}(e^{-i \beta}b)=\textrm{Im}(e^{-i \beta'}d)=0.\label{r1}
\end{align}
Moreover, assuming $c_{nm}$ is index independent and using the relation
\begin{align}
J_n(\alpha)J_m(\alpha') nm=&\frac{\alpha}{4}[J_{n+1}(\alpha)+J_{n-1}(\alpha)]J_m(\alpha') m\nonumber\\
+&\frac{\alpha'}{4}[J_{m+1}(\alpha')+J_{m-1}(\alpha')]J_n(\alpha) n,
\end{align}
Eq. (\ref{KGD3}) can be written as
\begin{align}\label{0alp}
&0=J_n(\alpha_m)J_m(\alpha'_n)r_1 n\nonumber\\
&+\frac{r_3m\alpha_m}{4}[J_{n-1}(\alpha_m)+J_{n+1}(\alpha_m)]J_m(\alpha'_n)\nonumber\\
&+\textrm{Re}(e^{-i\beta}b)[J_{n-1}(\alpha_m)J_m(\alpha'_{n-1})+J_{n+1}(\alpha_m)J_m(\alpha'_{n+1})]\nonumber\\
&+J_n(\alpha_m)J_m(\alpha'_n)r_2 m\nonumber\\
&+\frac{r_3n\alpha'_n}{4}[J_{m-1}(\alpha'_n)+J_{m+1}(\alpha'_n)]J_n(\alpha_m)\nonumber\\
&+\textrm{Re}(e^{-i\beta'}d)[J_{m-1}(\alpha'_n)J_n(\alpha_{m-1})+J_{m+1}(\alpha'_n)J_n(\alpha_{m+1})].
\end{align}
If we take the assumption that
\begin{align}\label{assump}
\alpha_{m\pm1}\approx\alpha_m\;\; \textrm{and}\;\; \alpha'_{n\pm1}\approx\alpha'_n,
\end{align}
Eq. (\ref{0alp}) can be considerably simplified and reduced into two equations
\begin{align}
&0=J_n(\alpha_m)r_1 n\nonumber\\
&+(\frac{r_3m\alpha_m}{4}+\textrm{Re}(e^{-i\beta}b))[J_{n-1}(\alpha_m)+J_{n+1}(\alpha_m)],\\
&0=J_m(\alpha'_n)r_2 m\nonumber\\
&+(\frac{r_3n\alpha'_n}{4}+\textrm{Re}(e^{-i\beta'}d))[J_{m-1}(\alpha'_n)+J_{m+1}(\alpha'_n)],
\end{align}
and thus it is straightforward to use the identity (\ref{BR}) to obtain
\begin{align}\label{alp}
&\alpha_m=-\frac{4\textrm{Re}(e^{-i\beta}b)}{2r_1+r_3 m},\\
&\alpha'_n=-\frac{4\textrm{Re}(e^{-i\beta}d)}{2r_2+r_3 n}.\label{alp1}
\end{align}
This result shows explicitly how the arguments of the Bessel functions depend on the indices and automatically explains the assumption in Eq. (\ref{Cnm}).

As a brief summary, the coefficients $c_{nm}$, $\beta$, $\beta'$, $\alpha_m$ and $\alpha'_n$ needed to determine the solution (\ref{psiG3}) can be obtained respectively from Eqs. (\ref{r0}, \ref{r1}, \ref{alp} and \ref{alp1}). The remaining task is to solve the Eq. (\ref{r0}) for the explicit form of $c_{nm}$ and finally check the validation of the assumption (\ref{assump}).

In the derivation for solving Eq. (\ref{r0}), we keep $\hbar$ explicitly for the benefit of indicating the perturbation orders. Presume the solution takes the form
\begin{equation}\label{cnm0}
c_{nm}=\eta e^{-\frac{i}{\hbar}[f(k\cdot x)+g(k'\cdot x)+q(k_d\cdot x)]},
\end{equation}
where $k_d=k-k'$ and $\eta$ is a constant; $f$, $g$ and $q$ are defined as real functions with the argument $k\cdot x$, $k'\cdot x$ and $k_d\cdot x$, respectively. Thus, there is
$\hat{p}_\mu c_{nm}=i\hbar\partial_\mu c_{nm}=c_{nm}[k_\mu f'+k'_\mu g'+(k_\mu-k'_\mu)q']$ with $f'$, $k'$ and $q'$ being the total differential of the corresponding functions with respect to the variables  $k\cdot x$, $k'\cdot x$ and $k_d\cdot x$, respectively.
Substituting this expression into  Eq. (\ref{r0}), it can be derived that
\begin{align}\label{cnm1}
k\cdot k'[&2f'g'+2(g'-f')q'-2q'^2-2i\hbar q''+\varrho_1\mp2\varrho_2g'\nonumber\\
&\mp2\varrho_3f'\mp2q'(\varrho_3-\varrho_2)]-2e^2aa'\cos(k_d\cdot x)\nonumber\\
&\mp2(f'k+g'k'+q'k_d)\cdot p_\pm=0,
\end{align}
with $\varrho_1=2\varrho_2\varrho_3$, $\varrho_2=e^2a^2/(2k\cdot p_\pm)$ and $\varrho_3=e^2a'^2/(2k'\cdot p_\pm)$.
Eq. (\ref{cnm1}) can be simplified by taking $f'=\pm\varrho_2$ and $g'=\pm\varrho_3$ for corresponding $p_\pm$, that
\begin{align}
k\cdot k'&(-2q'^2-2i\hbar q'')-2e^2aa'\cos(k_d\cdot x)\nonumber\\
&-2(\varrho_2 k+\varrho_3 k'\pm q'k_d)\cdot p_\pm=0,
\end{align}
Therefore,
\begin{align}\label{cnms}
&\pm q'k_d\cdot p_\pm+e^2aa'\cos(k_d\cdot x)+\frac{1}{2}e^2(a^2+a'^2)\nonumber\\
&=-(q'^2+i\hbar q'') k\cdot k'.
\end{align}

According to Eq. (\ref{cnms}) if $k_d\cdot p_\pm=0$, for example in a standing wave case $k'_z=-k_z$ and $a=a'$ with the particle beam shooting perpendicular to the $\hat{z}$ direction, the function
\begin{equation}
y(\phi)=e^{-\frac{i}{\hbar} q(2\phi)},
\end{equation}
with $\phi=k_d\cdot x/2$ satisfies the Mathieu differential equation
\begin{equation}\label{Mathi}
\frac{d^2y}{d\phi^2}+[c_1-2c_2\cos2\phi]y=0,
\end{equation}
with
\begin{equation}
c_1=-\frac{2e^2a^2}{\hbar^2 \omega^2},\; c_2=\frac{e^2a^2}{\hbar^2\omega^2}.
\end{equation}
A Mathieu equation is also found in solving the Klein-Gordon equation in a rotating electric field \cite{rotating}. The formal resemblance is reasonable since a particle locates in the vicinity of an antinode of the standing wave experiences a rotating electric field.

In the following we consider an oblique incidence of the particle into the laser field and suppose the condition
\begin{equation}\label{conlam}
|\lambda|=|\frac{k\cdot k'}{\pm k_d\cdot p_\pm}|\ll 1.
\end{equation}
Then Eq. (\ref{cnms}) can be solved by the perturbative method and the solution is obtained as a sum of terms proportional to different orders of $\lambda$, that
\begin{equation}\label{lambda}
q=q_{0}+\sum_{n=1}^{\infty} \lambda^n q_{n},
\end{equation}
where the zeroth-order solution takes a simple form
\begin{equation}
q_0=\mp\frac{e^2aa'\sin(k_d\cdot x)}{k_d\cdot p_\pm}\mp\frac{(\varrho_2 k+\varrho_3 k')\cdot p_\pm}{k_d\cdot p_\pm}k_d\cdot x .
\end{equation}
Therefore, the solution $c_{nm}$ reads
\begin{equation}\label{cnm2}
c_{nm}=\eta e^{\pm\frac{i}{\hbar}[-\varrho_2 k\cdot x-\varrho_3 k'\cdot x+\frac{e^2aa'\sin (k_d\cdot x)}{k_d\cdot p_\pm}+\frac{e^2(a^2+a'^2)}{2k_d\cdot p_\pm}k_d\cdot x+\mathcal{O}(\lambda)]}.
\end{equation}
Note that
\begin{equation} \label{dressm}
e^{\pm i p'_\pm\cdot x}c_{nm}=\eta e^{\pm i (p_\pm+\frac{e^2a^2+e^2a'^2}{2k_d\cdot p_\pm}k_d)\cdot x} e^{\pm i [\frac{e^2aa'\sin (k_d\cdot x)}{k_d\cdot p_\pm}+\mathcal{O}(\lambda)]},
\end{equation}
where the natural units $c=\hbar=1$ are used again.

The condition (\ref{conlam}) is satisfied in many scenarios. It means that there exists at least one inertial frame, here for the sake of simple illustration we assume it to be the lab frame, in which the particle asymptotic energy satisfies $\varepsilon\gg \omega(\omega')$ and we further require that $\varepsilon\gg m_e\xi (m_e\xi')$ with $\xi=\frac{ea}{m_e}$ and $\xi'=\frac{ea'}{m_e}$. This is referred to as the largest dynamical parameter condition in the paper similar to that in Ref. \cite{PiazzaG}.
Besides this energy requirement, the geometry of the relative movements of the particle and the laser waves can become significant in special cases, which will be discussed at the end of this section.
Then it can be calculated that
\begin{align}\label{r1app}
&r_1=\pm2k\cdot p_\pm(1\pm q'\frac{k\cdot k'}{k\cdot p_\pm})\approx \pm2k\cdot p_\pm,\\
&r_2=\pm2k'\cdot p_\pm(1\mp q'\frac{k\cdot k'}{k'\cdot p_\pm})\approx \pm2k'\cdot p_\pm,\label{r2app}
\end{align}
since the terms $q'\frac{k\cdot k'}{k\cdot p_\pm}, q'\frac{k\cdot k'}{k'\cdot p_\pm}\sim \frac{m_e^2(\xi^2+\xi'^2+\xi\xi')}{\varepsilon^2}$.
As stated earlier, the approximation presented in Eq. (\ref{assump}) needs to be checked. It is validated by
\begin{align}\label{dalp}
&|\alpha_{m\pm1}-\alpha_m|\approx\frac{m_e\xi\omega'\sin\theta_p}{\varepsilon\omega(1-\cos\theta_p)^2}\sim\frac{m_e\xi}{\varepsilon}\ll1,\\\label{dalp1}
&|\alpha'_{n\pm1}-\alpha'_n|\approx\frac{m_e\xi'\omega\sin\theta_p}{\varepsilon\omega'(1+\cos\theta_p)^2}\sim\frac{m_e\xi'}{\varepsilon}\ll1.
\end{align}

Therefore, the solution for the Klein-Gordon equation (\ref{KG3}) is obtained, which reads
\begin{align}\label{result3}
&\psi=\eta e^{\pm i (p_\pm+\frac{e^2a^2+e^2a'^2}{2k_d\cdot p_\pm}k_d)\cdot x} e^{\pm i [\frac{e^2aa'\sin (k_d\cdot x)}{k_d\cdot p_\pm}+\mathcal{O}(\lambda)]}\nonumber\\
&\times\sum_{n=-\infty}^{\infty}\sum_{m=-\infty}^{\infty}J_n(\alpha_m)e^{i n \beta}J_m(\alpha'_n)e^{i m \beta'}e^{i n k\cdot x}e^{i m k'\cdot x},
\end{align}
where
\begin{align}\label{result2}
&\cos\beta=\cos\beta'=\frac{p_\pm\cdot \epsilon_1}{\sqrt{(p_\pm\cdot \epsilon_1)^2+(p_\pm\cdot \epsilon_2)^2}},\\
&\sin\beta=\sin\beta'=-\frac{p_\pm\cdot \epsilon_2}{\sqrt{(p_\pm\cdot \epsilon_1)^2+(p_\pm\cdot \epsilon_2)^2}},\\\label{alpm}
&\alpha_m=-\frac{ea\sqrt{(p_\pm\cdot \epsilon_1)^2+(p_\pm\cdot \epsilon_2)^2}}{k\cdot p_\pm\pm\frac{m}{2}k\cdot k'},\\\label{alpn}
&\alpha'_n=-\frac{ea'\sqrt{(p_\pm\cdot \epsilon_1)^2+(p_\pm\cdot \epsilon_2)^2}}{k'\cdot p_\pm\pm\frac{n}{2}k\cdot k'},
\end{align}
and the constant $\eta$ is acquired in the normalization of the wavefunction, that $|\psi|^2=1$.

Finally the applicable range of the solution (\ref{result3}) is addressed. Firstly, let's scrutinize the steps (\ref{conlam}), (\ref{r1app}), (\ref{r2app}), (\ref{dalp}) and (\ref{dalp1}) which have taken approximations regarding four-vector multiplications. As mentioned previously, the validity of these approximations relies not only on the largest dynamical parameter condition, but also on the geometrical relations among the constituents' momenta. By checking the denominators of the expressions in these five equations, that specifically speaking $|k_d\cdot p_\pm|$, $|k\cdot p_\pm|$, and $|k'\cdot p_\pm|$, the conclusion is that: to justify these steps the azimuth angle $\theta_p$ of the particle's free momentum direction to the $\hat{z}$ axis should satisfy
\begin{equation}\label{con1}
\cos\theta_p\neq\frac{\omega-\omega'}{\omega+\omega'},
\end{equation}
and
\begin{equation}\label{con2}
\cos\theta_p\neq\pm1.
\end{equation}

Secondly, since the magnitudes of $\alpha_m$ and $\alpha'_n$ can be large since they scale as $m_e\xi/\omega$ and $m_e\xi'/\omega'$ respectively,  the usage of Eqs. (\ref{r1app}, \ref{r2app}) in Eqs. (\ref{alpm}, \ref{alpn}) can only ensure the neglected part is relatively much smaller than the remaining one. To guarantee the absolute magnitude of the neglected part to be $\ll 1$, extra constraints on the parameters are taken, that
\begin{equation}
|q'\frac{k\cdot k'}{k\cdot p_\pm}|\alpha_m\ll 1, \;\;\;|q'\frac{k\cdot k'}{k'\cdot p_\pm}|\alpha'_n\ll 1.
\end{equation}
These leads to the constraint on the field intensity
\begin{equation}\label{cons}
\xi\ll\frac{(\varepsilon^2\omega)^{\frac{1}{3}}}{m_e},\;\;\;\xi'\ll\frac{(\varepsilon^2\omega')^{\frac{1}{3}}}{m_e}.
\end{equation}

As an example, consider the two waves coming from the tunable Ti:Sa lasers with the photon energy $\omega=\omega'\sim2$eV and the particle being an electron accelerated by the laser-plasma wake field to the energy 10GeV \cite{10GeV} with $\theta_p=\pi/4$. Take $\xi\sim \xi'\sim 1$  which corresponds to the laser intensity about $10^{19}$W/cm$^{2}$. Then it can be calculated that $|\lambda|\sim10^{-10}$, $|q'\frac{k\cdot k'}{k\cdot p_\pm}|\sim |q'\frac{k\cdot k'}{k'\cdot p_\pm}|\sim 10^{-8}$, $|\alpha_{m\pm1}-\alpha_m|\sim|\alpha'_{n\pm1}-\alpha'_n|\sim10^{-4}$, and $|q'\frac{k\cdot k'}{k\cdot p_\pm}|\alpha_m\sim|q'\frac{k\cdot k'}{k'\cdot p_\pm}|\alpha'_n\sim10^{-2}$, and thus it is in the applicable range of the solution (\ref{result3}). Consider, as another example, the same particle condition while the waves are 1keV X-ray lasers with $\xi\sim \xi'\sim 1$ corresponding to the laser intensity about $10^{25}$W/cm$^2$ which is even higher than that available by the present technology, it can be found that $|\lambda|\sim10^{-7}$, $|q'\frac{k\cdot k'}{k\cdot p_\pm}|\sim |q'\frac{k\cdot k'}{k'\cdot p_\pm}|\sim 10^{-8}$, $|\alpha_{m\pm1}-\alpha_m|\sim|\alpha'_{n\pm1}-\alpha'_n|\sim10^{-4}$, and $|q'\frac{k\cdot k'}{k\cdot p_\pm}|\alpha_m\sim|q'\frac{k\cdot k'}{k'\cdot p_\pm}|\alpha'_n\sim10^{-5}$. Therefore it is also in the applicable range of the solution (\ref{result3}).


\section{IV. Recovering solutions in a plane-wave field}
By taking $A'=0$, it is found that the solution (\ref{result3}) is reduced to the familiar solution of Klein-Gordon equation in a plane-wave field as given in Eqs. (\ref{psiG}-\ref{result1}).

It is more interesting to note that by taking $k'$ in the same direction as $k$ and thus denoting $k'=rk$ with $r=\omega'/\omega$, the solution (\ref{result3}) becomes
\begin{align}\label{twocolor0}
&\psi=\eta e^{\pm i (p_\pm+\frac{e^2a^2}{2k\cdot p_\pm}k+\frac{e^2a'^2}{2k'\cdot p_\pm}k')\cdot x} e^{\pm i \frac{e^2aa'\sin (k_d\cdot x)}{k_d\cdot p_\pm}}\nonumber\\
&\times\sum_{n=-\infty}^{\infty}J_n(\alpha)e^{i n \beta}e^{i n k\cdot x}\sum_{m=-\infty}^{\infty}J_m(\alpha')e^{i m \beta'}e^{i m k'\cdot x},
\end{align}
where
\begin{align}
&\cos\beta=\cos\beta'=\frac{p_\pm\cdot \epsilon_1}{\sqrt{(p_\pm\cdot \epsilon_1)^2+(p_\pm\cdot \epsilon_2)^2}},\\
&\sin\beta=\sin\beta'=-\frac{p_\pm\cdot \epsilon_2}{\sqrt{(p_\pm\cdot \epsilon_1)^2+(p_\pm\cdot \epsilon_2)^2}},\\
&\alpha=-\frac{ea\sqrt{(p_\pm\cdot \epsilon_1)^2+(p_\pm\cdot \epsilon_2)^2}}{k\cdot p_\pm},\\
&\alpha'=-\frac{ea'\sqrt{(p_\pm\cdot \epsilon_1)^2+(p_\pm\cdot \epsilon_2)^2}}{k'\cdot p_\pm}.
\end{align}
Therefore
\begin{equation}\label{twocolor}
\psi=\eta e^{\pm i p_\pm\cdot x- i\int_{-\infty}^y dy'\frac{1}{2k\cdot p_\pm}(2ep_\pm\cdot A_t(y')\pm e^2A_t^2(y'))}
\end{equation}
with $y=k\cdot x$ and the total field potential $A_t(y)=A(y)+A'(ry)$, and thus the solution of Klein-Gordon equation in the two-color plane-wave field is recovered.

\section{V. Validation and simplification of the solution}
In the derivation for the solution, we have approximately solved Eq. (\ref{r0}) and obtained the arguments of the Bessel functions. They are the source of errors to the result. In Eq. (\ref{result3}), there is an error proportional to $\lambda\sim \omega/\varepsilon$ in the phase which can be diminished by calculating higher order terms. For the coefficient of each photon mode, the approximated arguments lead to an error proportional to $m_e\xi/\varepsilon$ or $m_e\xi'/\varepsilon$.

Let's consider the spectral width of the solution. The asymptotic formula of Bessel function for $n\gg \alpha>0$ gives
\begin{equation}
J_n(\alpha)\approx\frac{1}{\sqrt{2\pi n}}(\frac{e\alpha}{2n})^n,
\end{equation}
Thus the cutoff takes place at $n=\alpha$ and $J_n(\alpha)$ drops sharply as $n$ increases beyond $\alpha$ \cite{Erikthesis}.
For $J_n(\alpha)J_m(\alpha')$ with
\begin{align}
&\alpha=\frac{ea\sqrt{(p_\pm\cdot \epsilon_1)^2+(p_\pm\cdot \epsilon_2)^2}}{k\cdot p_\pm},\\
&\alpha'=\frac{ea'\sqrt{(p_\pm\cdot \epsilon_1)^2+(p_\pm\cdot \epsilon_2)^2}}{k'\cdot p_\pm},
\end{align}
the cutoff indices are $n_c=\pm \alpha$ and $m_c=\pm \alpha'$. These turn out to be approximately also the cutoff indices of the term $J_n(\alpha_m)J_m(\alpha'_n)$ in the solution (\ref{result3}) since there are $|\frac{n_c k\cdot k'}{k\cdot p_\pm}|\sim\frac{m_e\xi}{\varepsilon}\ll 1$ and $\frac{m_c k\cdot k'}{k'\cdot p_\pm}\sim\frac{m_e\xi'}{\varepsilon}\ll 1$.

To measure how well the approximated solution (\ref{result3}) satisfies the Klein-Gordon equation (\ref{KG3}), let's calculate the expectation value of the left-hand side of Eq. (\ref{KG3}) with Dirac bra-ket notation
\begin{equation}
\delta \varepsilon^2=<\psi|(\hat{p}-eA-eA')^2-m_e^2|\psi>.
\end{equation}
This is in fact the temporal-spatial integration of the multiplication of the conjugate of the solution (\ref{result3}) and the right-hand side of Eq. (\ref{0alp}). It is reasonable to compare this residual energy with the characteristic energy of the particle, for example $\varepsilon^2$ in the lab frame or Lorentz invariants like $k\cdot p_\pm$. Without going into the details and focusing only on the energy scales, we get
\begin{equation}\label{con3}
\frac{\delta\varepsilon^2}{\varepsilon^2}\sim \frac{m_e^2\xi\xi'n_cm_c}{\varepsilon^2}\sim \frac{(m_e^2\xi\xi')^2}{\omega\omega'\varepsilon^2}.
\end{equation}
In the Ti:Sa laser example given at the end of section 2, $\frac{(m^2_e\xi\xi')^2}{\omega\omega'\varepsilon^2}\approx 10^2$. To make the quantity small, one way is to reduce the intensity of the laser field to be, e.g., $\xi\sim\xi'\sim 0.1$ corresponding to the intensity about $10^{17}$W/cm$^2$, and $\frac{(m^2_e\xi\xi')^2}{\omega\omega'\varepsilon^2}\approx 10^{-2}$. In the X-ray example also at the end of section 2, there is $\frac{(m^2_e\xi\xi')^2}{\omega\omega'\varepsilon^2}\approx 10^{-2}$.

It is tempting to write the solution (\ref{result3}) in a more concise form via using the identity
\begin{equation}\label{iden}
e^{ix\sin\theta}=\sum_{-\infty}^{\infty}J_n(x)e^{in\theta},
\end{equation}
like what has been done to the solution (\ref{twocolor0}) and (\ref{twocolor}).
But due to the index-dependent arguments $\alpha_m$ and $\alpha'_n$ of the Bessel functions in the solution (\ref{result3}), not only the direct application of this identity is incorrect, but also the sum there can not be decomposed as a product of two sums like in Eq. (\ref{twocolor0}). This illustrates the complex manner of the particle coupling with the non-plane-wave field composed of two counter propagating laser waves.

However, if at the cutoff there is
\begin{align}\label{str1}
|\alpha_{m_c}-\alpha|\sim |\alpha\alpha'\frac{\omega'}{\varepsilon}|\sim \frac{m^2_e\xi\xi'}{\omega\varepsilon}\ll 1,\\\label{str2}
|\alpha'_{n_c}-\alpha'|\sim |\alpha\alpha'\frac{\omega}{\varepsilon}|\sim \frac{m^2_e\xi\xi'}{\omega'\varepsilon}\ll 1,
\end{align}
which compared to Eq. (\ref{con3}) just means the energy scale of $\delta \varepsilon/\varepsilon$ is small, then the following approximation is reasonable
\begin{equation}
\alpha_{m}\approx\alpha,\;\;\; \alpha'_{n}\approx\alpha'.
\end{equation}
Therefore the solution is simplified as
\begin{equation}
\psi=e^{iS},
\end{equation}
where the action function reads
\begin{align}\label{S}
S=&\pm (p_\pm+\frac{e^2a^2+e^2a'^2}{2k_d\cdot p_\pm}k_d)\cdot x\pm\frac{e^2aa'\sin(k_d\cdot x)}{k_d\cdot p_\pm}\nonumber\\
&-[\frac{eap_\pm\cdot \epsilon_1}{k\cdot p_\pm}\sin(k\cdot x)-\frac{eap_\pm\cdot \epsilon_2}{k\cdot p_\pm}\cos(k\cdot x)]\nonumber\\
&-[\frac{ea'p_\pm\cdot \epsilon_1}{k'\cdot p_\pm}\sin(k'\cdot x)-\frac{ea'p_\pm\cdot \epsilon_2}{k'\cdot p_\pm}\cos(k'\cdot x)],
\end{align}
or in an integral form
\begin{align}\label{Sint}
S=&\pm (p_\pm+\frac{e^2a^2+e^2a'^2}{2k_d\cdot p_\pm}k_d)\cdot x\pm\int_{-\infty}^y dy'\frac{1}{k_d\cdot p_\pm}e^2 F(y')\nonumber\\
&- \int_{-\infty}^y dy'\frac{1}{k\cdot p_\pm}ep_\pm\cdot A(y')- \int_{-\infty}^y dy'\frac{1}{k'\cdot p_\pm}ep_\pm\cdot A'(y').
\end{align}
where $F(y)=-A\cdot A'$ with $y=k_d\cdot x$. In analogy to the plane-wave solution, the dressed momentum of the particle in this non-plane-wave field can be identified as
\begin{equation}
q=p_\pm+\frac{e^2a^2+e^2a'^2}{2k_d\cdot p_\pm}k_d,
\end{equation}
and accordingly the dressed mass is
\begin{equation}
m_*=\sqrt{q^2}\approx m_e\sqrt{1+\xi^2+\xi'^2}.
\end{equation}

The validity of action (\ref{S}) can be checked by substituting it into Eq. (\ref{action}). Then the left-hand side of the equation reads
\begin{align}
&(\partial_\mu S+eA_{\mu}+eA'_\mu)(\partial^\mu S+eA^\mu+eA'^\mu)-m_e^2 \nonumber\\
\sim&\frac{(m_e\xi)^3}{\varepsilon}+m_e^2\xi\xi',
\end{align}
and with $\frac{m_e\xi^3}{\varepsilon}\ll1$ it is
\begin{equation}\label{result4}
(\partial_\mu S+eA_\mu+eA'_\mu)(\partial^\mu S+eA^\mu+eA'^\mu)-m_e^2(1-\mathcal{O}(\xi\xi'))=0.
\end{equation}
Like above, the ratio of the extra term over the characteristic energy of the particle $\sim m^2_e\xi\xi'/\varepsilon^2\ll 1$. In fact there are meaningful strong-field problems with $\xi\sim 1$ and $\xi'\ll 1$ which results in $\xi\xi'\ll 1$.

In fact, it is possible to construct other forms of expressions which can also make the Eq. (\ref{action}) established up to a certain perturbation order, but this paper shows how this particular form (\ref{S}) of the action function is derived step by step and proves its reasonableness. It takes the form analogous to that of the plane-wave solution and laser-dressed physical quantities such as the dressed momentum and the dressed mass can be directly identified.

It is worth noticing that neglecting in Eq. (\ref{0alp}) the terms with $r_3m\alpha_m$ and $r_3n\alpha'_n$, the arguments of the Bessel functions would be index independent and the action function (\ref{S}) can be obtained. However, if the error brought to the Klein-Gordon equation is estimated based on Eq. (\ref{0alp}), an energy scale of $\frac{m_e^3\xi^2\xi'}{\omega}$ or $\frac{m_e^3\xi\xi'^2}{\omega'}$ is found which is more severe than $m_e^2\xi\xi'$ found in Eq. (\ref{result4}) by direct calculation with the explicit expression of the action function.


\section{VI. Conclusions}
The analytical solution of the Klein-Gordon equation and equivalently the action function of a Dirac particle in a non-plane-wave electromagnetic field are investigated. The method is developed based on the idea that the coefficients of the Fourier expansion terms of the solution to different photon modes mainly adopt the form of multiplications of Bessel functions.

For the field composed of two counter propagating circularly polarized plane waves, a detailed derivation is illustrated. The coefficients are determined by the rules (\ref{r0}), (\ref{r1}), (\ref{alp}) and (\ref{alp1}) as a development or direct analogy to the rules observed in the plane wave case, see Eqs. (\ref{rule1}), (\ref{rule2}) and (\ref{rule3}). In order to solve Eq. (\ref{r0}) explicitly and justify the assumption (\ref{assump}), the largest dynamical parameter condition is imposed, that there exists an inertial frame in which the particle free momentum is far larger than the other field dynamical parameters.  As already mentioned in section 1, this condition is of realistic meaning in view of strong-field experimental campaigns into quantum nonlinearity regime. The solution for the Klein-Gordon equation is obtained analytically, see Eqs. (\ref{result3}---\ref{alpn}). Discussions on its applicable range and examples can be found at the end of section 3.

It is clearly shown in section 4 that the solutions for the Klein-Gordon equation in the one-color and two-color plane-wave fields can be recovered. Considering the cutoff property of the Bessel function, it is found in section 5 that the solution can be simplified and the action function takes an integral form (\ref{Sint}) similar to that in the plane-wave case. The laser-dressed momentum and mass of the particle are identified. The validity of the simplified action is justified by directly calculating the basic equation (\ref{action}) that defines the action function. Comparing the simplified action (\ref{S}) and Eq. (\ref{cnm2}), it can be found that the non-plane-wave feature of the coupling of the particle to the two waves is mainly determined by Eq. (\ref{r0}).

The solution (\ref{result3}---\ref{alpn}) as well as the action (\ref{Sint}) is Lorentz invariant. It has the advantage of clear structure and close similarity to the solution in a plane-wave case, and thus favors convenient application. With the solution of the Klein-Gordon equation, the solution of the corresponding Dirac equation in the same electromagnetic field can be derived by the method shown in paper \cite{PiazzaG}. It then can provide more exact reaction rates of multi-photon pair production process, Compton scattering process and so on in this non-plane-wave field. By comparing the results calculated from the plane-wave solution and the non-plane-wave solution, novel features particularly related to the non-plane-wave field can be identified and used in experimental design and explanation.

This paper indicates the special convenience of using Bessel functions in describing the dressed states of particles in electromagnetic fields. Extending this method to solve problems in other field configurations shall be investigated in the future work.

We are grateful to Dr. B. King and Prof. H. R. Reiss for fruitful discussions and valuable advice. This work is financially supported by the National Natural Science Foundation of China under Grant No. 11204370.

\end{document}